# Weak antilocalization of ultrarelativistic fermions

K.Yu. Bliokh[1,2*]

[1]*Institute of Radio Astronomy, 4 Krasnoznamyonnaya st., Kharkov, 61002, Ukraine*
[2]*Department of Physics, Bar-Ilan University, Ramat Gan, 52900, Israel*

The paper discusses the Berry phase influence on the weak localization phenomenon at the adiabatic backscattering of ultrarelativistic particles in a random medium. We demonstrate that bosons that pass along a certain closed path in opposite directions come back always in phase (an example: the backscattering enhancement of electromagnetic waves), whereas fermions come exactly in antiphase. This produces a "complete weak antilocalization" of ultrarelativistic fermions: the backscattering field intensity vanishes.

The phenomenon of backscattering enhancement has been much studied both for electromagnetic waves (photons), which are scattered in randomly inhomogeneous media (for review see [1]), and for electrons in solids (for review see [2]). In the latter case, the effect is commonly referred to as weak localization of quantum particles. The essence of the phenomenon consists in the following. In the course of multiple scattering of a wave in a random medium at an arbitrary (different from $\pi$) finite angle, the partial waves traveling along various trajectories come to an observation point with random phases. As a result, an *incoherent* summation takes place, and the total intensity of scattered field is determined as a sum of the partial intensities. On the contrary, in the case of the backscattering, i.e. scattering at $\pi$ angle, all the trajectories are loop-shaped, and any chosen trajectory can be associated with the same trajectory passed in opposite direction (Fig. 1). As a result, all the trajectories are grouped in pairs, and the phase increments coincide for both directions of the loop passage. Hence a *coherent* summation of fields occurs, and one should sum the field amplitudes from the paired trajectories. It is easy to see that in this case the backscattering intensity is exactly twice as large as the intensity of the field scattered in other directions [1] (the scattering indicatrix is considered to be isotropic) (see Fig. 3 below).

Incorporation of various interactions can essentially change the features of the backscattering enhancement. If the phase difference of the waves traveling along a loop-trajectories in opposite directions changes, the backscattering enhancement can be reduced, vanish or even be replaced by attenuation [2,3]. In the latter case, the *weak antilocalization* of particles is said to occur. For electrons, two basic mechanisms for the interaction's influence on the weak localization are known. The first one is associated with the introduction of an external magnetic field. If a magnetic flux $\Phi$ penetrates a loop-trajectory, the electrons travelling this loop in opposite directions acquire the Dirac phase difference $2e\Phi/\hbar c$ (the Aharonov-Bohm effect). Due to a random character of the trajectories, the flux $\Phi$, as well as the phase difference, turn out to be random values also. As a consequence of this, the backscattering peak reduces as the magnetic field increases, and it disappears when the magnetic field is strong enough: the backscattered fields become incoherent. In solids, the weakening of the electron backscattering leads to the growth of the conductivity as the magnetic field increases, or in other words, to the *negative magnetoresistance (magnetoconductance)* effect [2,3]. Another mechanism of the change in the phase difference on the paired trajectories is the appearance of the Berry geometric phase. Berry phase for electrons arises owing to the presence of a spin and is connected to the Zeeman and spin-orbit interactions. Therefore, the weak electron antilocalization is frequently associated with the spin-orbit interaction. In contrast to Dirac phase, Berry phase does not depend explicitly on the loop shape and is determined only by the geometry of the particle motion in the momentum space. Due to this, the Berry phase can be a regular quantity even for random loop-trajectories (see below). Then the backscattered fields are summed coherently as before, but the Berry phase's difference may cause both the enhancement and weakening of the backscattered field, i.e., it may cause the weak antilocalization [2,3].

---
[*]E-mail: k_bliokh@mail.ru

The weak antilocalization is not observed in the electromagnetic wave scattering [4]. Meanwhile, electromagnetic waves consist of photons, i.e. relativistic particles with spin 1. They also possess the Berry geometric phase [5]; moreover, as it was recently shown, the Berry phase can be presented as a consequence of the spin-orbit interaction of photons [6–10]. Why does the presence of the Berry phase not result in the weak antilocalization in photons? Is the weak antilocalization effect possible for other ultrarelativistic particles?

First, let us recall that in the case of adiabatic evolution (if the helicity is conserved), the Berry phase for photons and other ultrarelativistic particles (i.e. a particle with the energy much larger than the rest energy) of arbitrary spin $s$ is determined by a simple geometric law [5,11]. Let us consider a closed trajectory of a particle motion in the momentum $\mathbf{p}$-space. The presence of spin freedom degrees in the relativistic wave equations entails that the particle moves as if an effective 'magnetic monopole' of charge $\sigma$ (where $\sigma = -s, -s+1, ..., s-1, s$ is the helicity of the particle and $\sigma \neq 0$ for photons) is placed at the origin of $\mathbf{p}$-space [5,8,10,11]. As a result, the Berry phase $\vartheta_B$ is merely an analogue of the Dirac phase generated by this monopole; it is numerically equal to the flux passing through the contour of the particle trajectory, or

$$\theta_B = \sigma \Omega \ , \tag{1}$$

where $\Omega$ is the solid angle at which the trajectory loop is visible from the origin of the $\mathbf{p}$-space. Owing to this law, a particle motion in the $\mathbf{p}$-space is conveniently projected on a unit sphere (i.e. a space of vectors $\mathbf{t} = \mathbf{p}/|\mathbf{p}|$: the tangents to a trajectory in the coordinate $\mathbf{r}$-space), on which the area bounded by the loop is equal to $\Omega$.

Now let us consider a trajectory of an ultrarelativistic particle adiabatically backscattered due to multiple scatterings in a random medium. (As is shown for photons, in certain situations, the helicity is conserved for long enough distances, which allows one to use the adiabatic approximation [12]). Fig. 2 shows typical particle trajectories on the unit sphere in the $\mathbf{p}$-space. Trajectories '1' and '2' correspond to the passage along the same $\mathbf{r}$-space loop in opposite directions (Fig. 1); they are symmetric with respect to the center of the sphere. The Berry phase difference for the wave fields that have passed along these two trajectories is given by formula (1), where $\Omega$ is the area on the sphere bounded by the trajectories. Evidently, $\Omega = 2\pi$, or $\Omega = 2\pi + 4\pi k$ ($k \in \mathbb{Z}$) in the general case, and equation (1) yields

$$\theta_B = 2\pi\sigma + 4\pi k\sigma \ . \tag{2}$$

For bosons, $\sigma$ is an integer, while for fermions it is a half-integer value, and we derive from (2) that

$$\theta_B^{(Bose)} = 0 \mod 2\pi \ , \quad \theta_B^{(Fermi)} = \pi \mod 2\pi \ . \tag{3}$$

This result implies that *when backscattering occurs in a random medium, the ultrarelativistic bosons that pass along the same loop trajectory in opposite directions come back always in phase, whereas the fermions arrive in antiphase.* Hence, *a "complete weak antilocalization" occurs for ultrarelativistic fermions.* According to (3), the Berry phase does not affect a weak localization of electromagnetic waves since they represent bosons. A magnitude of the Berry phase difference is insensitive to the trajectory details and may be thought of as constant within the limits of a narrow backscattering peak. Therefore the dependence of the intensity $I$ on the scattering angle $\varphi$ for fermions is almost identical to that for bosons, but with the backscattering peak 'inverted' (Fig. 3). In particular, *when scattered at angle $\pi$, the field intensity vanishes.* One can write that

$$I(\varphi) \simeq I_0 \pm \delta I(\varphi) ,  \qquad (4)$$

where $I_0$ is the isotropic background of the incoherent scattering, $\delta I(\varphi)$ stands for the coherent corrections with $\delta I(\pi) = I_0$ and $\delta I \simeq 0$ everywhere except a narrow band $|\varphi - \pi| \sim \lambda/L$ ($\lambda$ is the wavelength and $L$ is the typical scale of scattering) [1], while the signs '+' and '−' correspond to bosons and fermions respectively.

The results obtained are in qualitative accordance with the results for nonrelativistic electrons in solids with a spin-orbit interaction [2,3]. However, it should be emphasized that for nonrelativistic electrons the influence of the spin-orbit interaction and Berry phase on the localization is of a competitive character and depends on the strength of the spin-orbit interaction, while in the case of ultrarelativistic particles the Berry phase drastically changes the picture to its opposite. The introduction of the magnetic field in the scattering of fermions of a non-zero charge is bound to cause the addition of a random Dirac phase difference for the particles that have travelled along the paired trajectories. As a result, the backscattering intensity will increase up to the isotropic level in the strong-field limit. An experimental discovery of the considered phenomenon faces a number of obstacles associated with the narrowness of the weak antilocaliztion peak and the provision of a multiple adiabatic scattering. Possibly, the effect can be observed during the scattering of relativistic electrons by complex molecules, where the multiple scattering from the different components of the molecule can occur. (The small distances between scatterers offer greater thickness of the peak.) Even with an anisotropic scattering indicatrix, a coherent component will cause the weakening of the field intensity in strictly backward direction. Besides, a quite similar to above considered situation can occur in condensed matter physics, if the carriers demonstrate the Berry gauge field of the magnetic monopole type and the Berry phase similar to (1) with an odd helicity. In particular, such situation takes place in hole-doped semiconductors described by Luttinger Hamiltonian with $\sigma = \pm 3/2$ and $\pm 1/2$ for heavy-hole and light-hole bands, respectively [13]. In any case, the presented theoretical results demonstrate that geometry and statistics can play a crucial role in the phenomenon of weak (anti)localization of quantum particles with a spin.

This work has been partially supported by INTAS (Grant No. 03-55-1921) and Ukrainian President's Grant for Young Scientists GP/F8/51.

**FIGURE CAPTIONS**

**Fig. 1.** Paired loop trajectories at the backscattering.

**Fig. 2.** Trajectories on the unit sphere in the momentum space, which correspond to the paired loop trajectories of Fig. 1. The 'north' and the 'south' poles of the sphere correspond to the directions of incident and backscattered waves respectively.

**Fig. 3.** The scattering indicatrix in a random medium: the dashed peak corresponds to the usual weak localization (the boson case), while the solid line corresponds to the "complete weak antilocalization" (the fermion case). The arrow points the wave incidence direction.

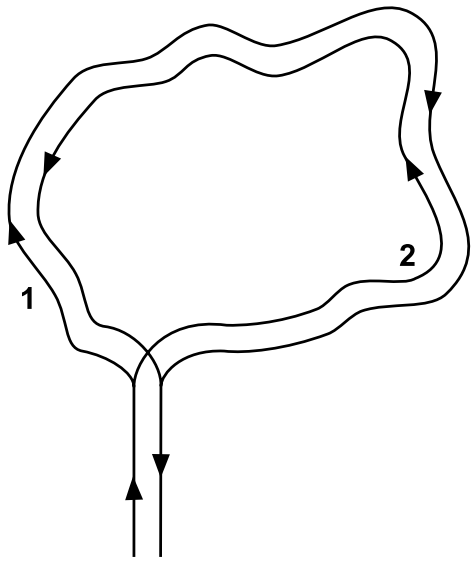

**Fig. 1**

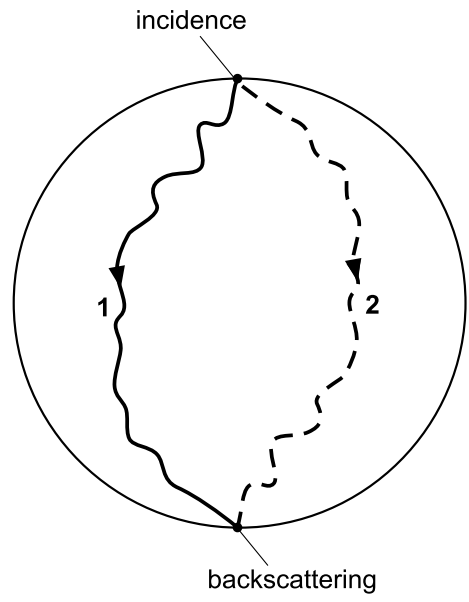

**Fig. 2**

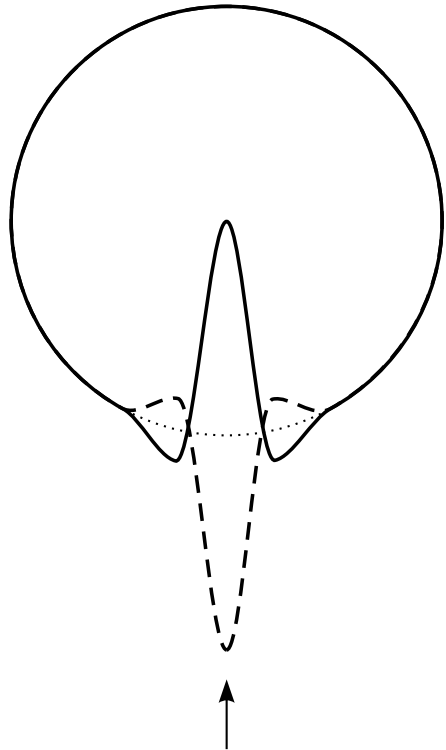

**Fig. 3**